\documentclass[10pt,journal,compsoc]{IEEEtran}
\ifCLASSOPTIONcompsoc
  \usepackage[nocompress]{cite}
\else
  \usepackage{cite}
\fi
\usepackage{amssymb}

\ifCLASSINFOpdf
   \usepackage[pdftex]{graphicx}
\else
\fi
\usepackage{amsmath}
\usepackage{gensymb}
\usepackage{colortbl}
\usepackage{svg}
\usepackage{multirow}
\usepackage{acronym}
\newacro{CNN}[CNN]{Convolutional Neural Network}
\newacro{CCM}[CCM]{Color Conversion Matrix}
\newacro{RAE}[RAE]{recovery angular error}
\newacro{C3AE}[C3AE]{Color Constancy Convolutional AutoEncoder}
\newacro{HVS}[HVS]{Human Visual System}
\newacro{NLP}[NLP]{Natural Language Processing}
\newacro{GAN}[GAN]{Generative Adversarial Network}
\newacro{BoCF}[BoCF]{Bag of Color Features}
\newacro{ReLU}[ReLU]{Rectified Linear Units}
\newacro{GDPR}[GDPR]{General Data Protection Regulation}

\makeatletter
\renewcommand\@makefntext[1]{\leftskip=2em\hskip-2em\@makefnmark#1}
\makeatother

\begin{document}
%
\title{INTEL-TAU: A Color Constancy Dataset}

%
%

\author{Firas~Laakom,
Jenni~Raitoharju,~\IEEEmembership{Member,~IEEE,}
Jarno~Nikkanen,~\IEEEmembership{Member,~IEEE,}
Alexandros~Iosifidis,~\IEEEmembership{Senior~Member,~IEEE,} and~Moncef~Gabbouj,~\IEEEmembership{Fellow,~IEEE}


       \thanks{F. Laakom, J. Raitoharju,  and M. Gabbouj are with Faculty of Information Technology and Communication Sciences, Tampere University, Tampere, Finland (e-mail: firas.laakom@tuni.fi; jenni.raitoharju@tuni.fi; moncef.gabbouj@tuni.fi).}
       \thanks{A. Iosifidis is with the Department of Engineering, Aarhus University, DK-8200 Aarhus, Denmark (e-mail: ai@eng.au.dk).}
       \thanks{J. Nikkanen is with INTEL, Insin\"{o}\"{o}rinkatu 41, 33720 Tampere, Finland (email: jarno.nikkanen@intel.com) }
}
\IEEEtitleabstractindextext{%
\begin{abstract}
In this paper, we describe a new large dataset for illumination estimation. This dataset, called INTEL-TAU, contains 7022 images in total, which makes it the largest available high-resolution dataset for illumination estimation research. The variety of scenes captured using three different camera models, namely Canon 5DSR, Nikon D810, and Sony IMX135, makes the dataset appropriate for  evaluating the camera and scene invariance of the different illumination estimation techniques. Privacy masking is done for sensitive information, e.g., faces. Thus, the dataset is coherent with the new \ac{GDPR}. Furthermore, the effect of color shading for mobile images can be evaluated with INTEL-TAU dataset, as both corrected and uncorrected versions of the raw data are provided. Furthermore, this paper benchmarks several color constancy approaches on the proposed dataset.  
\end{abstract}

}

\maketitle

\IEEEdisplaynontitleabstractindextext

%
\IEEEpeerreviewmaketitle

\IEEEraisesectionheading{\section{Introduction}\label{sec:introduction}}

\IEEEPARstart{T}{he} observed color of an object in a scene depends on its spectral reflectance and  spectral composition of the illuminant. As a result, when the scene illuminant changes, the light reflected from the object also changes \cite{ebner2007color,8410422}.  The ability to filter out the color of the light source is called color constancy \cite{ebner2007color,logvinenko2015rethinking}. executing this ability is critical for many image processing and computer vision applications. It results in better quality images. For a robust color-based system, the illumination effects of the light source need to be discounted, so that colors present in the image reflect the intrinsic properties of the objects in the scene \cite{8410422,halimi2016fast,kellman2019physics}. This is important for many high level image or video applications. Without computational color constancy, colors would be an unreliable feature and inconsistent for object recognition, detection, and tracking. Thus, color constancy research, also called illumination estimation, has been extensively studied and several approaches have been proposed to tackle it\cite{3, 46440,ratina, chzcc2011,afifi2020deep,afifi2019sensor}.

One key assumption of classical computational color constancy approaches is that the illumination of a scene is uniform.  Thus, the problem can be divided into two main steps. In the first step, the global illumination is estimated and, in the second step, all color pixels of the scene are normalized using the estimated illuminant color.  As the second step is a straight-forward transformation, the computational color constancy problem is equivalent to illumination estimation.  Typically, illumination estimation algorithms are divided into two main groups, namely unsupervised approaches and supervised approaches. The former involves methods requiring no training which are based on low-level statistics \cite{yang2015efficient,d5,d8,d6,laakom2020probabilistic}  and methods using physics-based dichromatic reflection model \cite{3,8,11,LuICCV2009}, while the latter involves data-driven approaches that learn to estimate the illuminant in a supervised manner using labeled data.

With the advancement of machine learning in general and deep learning in particular, many machine learning-based approaches have been proposed for color constancy \cite{7298702,4664624, 22, 44, DSN, Barron2015ConvolutionalCC, 34, f1, mine}. However, machine learning-based approaches, especially methods relying on \acp{CNN}, usually have a large number of parameters that need to be optimized for solving the illumination estimation problem.   Thus, training such models require a large amount of labeled data for training and evaluation. Moreover, The performance of such methods in the test scenario heavily depends on the quality and the diversity of the data seen during the training process. However, acquiring labeled datasets for illumination estimation is a challenging task \cite{Hemrit2018RehabilitatingTC}, as in order to extract the ground truth illumination of a scene, a ColorChecker chart needs to be included in the scene. In addition, after the introduction of \acf{GDPR} act \cite{voigt2017eu} in Europe, data privacy in datasets needs to be addressed and sensitive information needs to be masked.

\begin{figure}[tb]
\centering
\includegraphics[width= 0.5\textwidth]{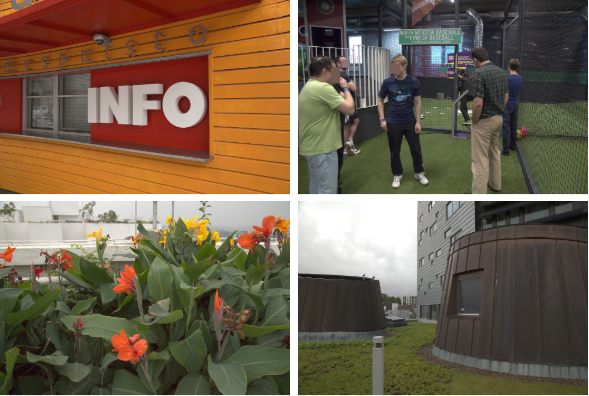}
\caption{Samples from INTEL-TAU dataset}
\label{figsamples}
\end{figure}

\begin{table*}[h]

\renewcommand{\arraystretch}{1}
 \centering	
	\caption{Characteristics of different high-resolution color constancy datasets}
	\label{tab:parameters}
	\begin{tabular}{l|ccccccc} 
		\hline
Dataset & ColorChecker (Gehler's Raw)  & Cube  & cube+  & SFU HDR & NUS-8 & INTEL-TUT & \textbf{INTEL-TAU}\\	
\hline
Number of samples  &  568 & 1365  & 1707 & 105  & 1736  &  1558 & \textbf{7022} \\
Number of camera models & 2 & 1  & 1  & 1 & 8 & 3 & \textbf{3} \\
Indoor and outdoor samples & \checkmark & --  & \checkmark  & \checkmark  & \checkmark  & \checkmark  & \checkmark      \\
\ac{GDPR} compliance     &  -- &  -- &  -- &  --  &  -- &  --  & \checkmark \\
Publicly available  & \checkmark  & \checkmark & \checkmark  & \checkmark & \checkmark  & -- & \checkmark  \\
\end{tabular}
\label{tab:labeltable1}
\end{table*}

In this paper, we propose a new INTEL-TAU dataset for color constancy research. The dataset contains 7022 high-resolution images and it is by far the largest publicly available high-resolution dataset for training and evaluation of color constancy algorithms. Furthermore, all recognizable faces, license plates, and other privacy sensitive information have been carefully masked. Thus, the dataset is now fully \ac{GDPR} compliant. A subset of 1558 images of the current dataset was previously published as Intel-TUT dataset \cite{17}, but had to be retracted due to its \ac{GDPR} non-compliance. Images in INTEL-TAU dataset were collected  using three different cameras: Canon 5DSR, Nikon D810, and Mobile Sony IMX135. The images contain both field and lab scenes. The dataset has mainly real scenes along with some lab printouts with the corresponding white point information. The black level was subtracted from all the scenes and the saturation points, i.e., where the incident light at a pixel causes one of the color channels of the camera sensor to respond at its maximum value producing an undesirable artifact, were normalized. This dataset is suitable for scene and camera-invariance estimation of color constancy algorithms.

The rest of this paper is organized as follows. First, we review the available color constancy datasets  in Section \ref{sec:relatedwork}. In Section \ref{sec:datasetdescription}, we describe the proposed dataset and highlight its main novelties.  We propose several protocols for using this dataset for illumination estimation research in Section \ref{exprotocols}. In Section \ref{sec:experimentalresults}, we evaluate the performance of several baseline and recently proposed color constancy algorithms on the proposed dataset. We conclude the paper in Section \ref{sec:conclusion}.

\section{Previously published color constancy datasets} \label{sec:relatedwork}

One of the most commonly used dataset in color constancy is the  ColorChecker dataset \cite{47}. It is composed of 568 high-resolution raw images acquired by two cameras: Canon 1D and Canon 5D. Shi and Funt \cite{shire} proposed a methodology to reprocess the original images and to recalculate the ground truth.  The images are demosaiced and available as TIFF images. The location of the color chart and the saturated and clipped pixels are also provided with the database. Later, Finlayson et al. \cite{finlayson2017curious} raised a problem, regarding the computation of the ground truth, with the Shi reprocessed dataset. To remedy this problem, a Recommended ColorChecker dataset with an updated ground truth was introduced in \cite{21,Hemrit2018RehabilitatingTC}. 

Another publicly available dataset is SFU HDR \cite{MaxRGB,rehabilitation} containing 105 high dynamic range images captured using a calibrated camera. Nine images per scene were captured in order to generate the high dynamic range images. For an accurate measure of  global illumination, four color charts were used at different locations in the scene.

NUS-8 \cite{nus} has been one of the largest color constancy datasets. It contains 1736 raw images. Eight different camera models were used to capture the scenes of this dataset and a total of $\sim210$ images were captured by each camera.  Although the dataset is relatively large, a commonly used protocol is to perform tests on each camera separately and report the mean of all the results. As a result, each experiment involves using only 210 images for both training and testing, which is not enough to appropriately train deep learning-based approaches.

Banic and Loncaric introduced the  Cube dataset in \cite{banic2017unsupervised}. This dataset is composed of 1365 RGB images. All dataset images are outdoor scenes acquired with a Canon EOS 550D camera in Croatia, Slovenia, and Austria. This dataset was also extended into Cube+ dataset  \cite{banic2017unsupervised}. This extension was enriched by an additional 342 images containing indoor and outdoor scenes. The overall distribution of illuminations in the Cube+ is similar to the ground truth distribution of the NUS-8.

Other hyperspectral datasets \cite{hyperspectral,hyperspectral2,hyperspectral3,ciurea2003large} are available for color constancy research. However, these dataset are relatively scarce and thus unsuitable for machine learning-based solutions with the exception of \cite{ciurea2003large} which contains 11000 images. However, this dataset is actually composed of video frames and, as a result, most of the images are highly correlated and only 600 are not \cite{47}. Moreover, this dataset has low-resolution images that were subject to correction.

A different research direction in computational  color constancy is the multi-frame illumination estimation \cite{qian2017recurrent,beigpour2016multi}.  Several  video-based  datasets  have  been  proposed  to incorporate  the  temporal  information  in  the  learning  process \cite{qian2020benchmark, prinet2013illuminant, yoo2019dichromatic}. Another noteworthy type of datasets are the multi-illumination datasets \cite{murmann2019dataset, aksoy2018dataset, hui2016white,hao2019evaluating}. It has been argued that many inverse problems involving lighting and material understanding remain severely ill-posed to be solved efficiently with single-illumination datasets \cite{murmann2019dataset,afifi2019color,afifi2019else}. However, capturing labeled samples of multiple illuminants is hard and, thus, the available datasets are typically small and collected in a controlled environment \cite{murmann2019dataset}, which make them  impractical for real-world applications.  Thus, there is still a need for larger single-illumination datasets.

Intel-TUT was proposed in \cite{17}. It contained a subset of 1558 images of the proposed INTEL-TAU dataset. Due to the aforementioned problems with \ac{GDPR} regulations, it was recently retracted. Furthermore, a larger subset of 3420 images was  used for experiments in \cite{f1} and \cite{mine}, 
but similar privacy issues were encountered. The privacy masking, which we applied for the proposed INTEL-TAU dataset, resolves all the privacy issues, while preserving all the advantages of the previously published subsets and providing further benefits as described in the next section.  Table \ref{tab:labeltable1} presents a comparison of different color constancy datasets.

\section{INTEL-TAU Dataset description} \label{sec:datasetdescription}

We introduce a new color constancy dataset, called INTEL-TAU, with the following properties. INTEL-TAU
\begin{itemize}
    \item is  currently the largest publicly available high-resolution  color constancy dataset containing 7022 images  with ground truth illumination,
    \item is available at http://urn.fi/urn:nbn:fi:att:f8b62270-d471-4036-b427-f21bce32b965, 
    \item  provides the training images without the color charts (i.e., there is no need for color chart masking),
    \item contains images taken with 3 different cameras to allow camera invariance evaluation,
    \item contains images grouped by  scene type to allow scene invariance evaluation,
    \item contains mobile images before and after color shading\footnote{The non-shaded mobile images are available in the additional resources discussed in Section \ref{addit}} to allow studying the effect of shading,
    \item is fully \ac{GDPR} compliant with privacy masking applied on all sensitive information,   
  
\end{itemize}


INTEL-TAU contains both outdoor and indoor images captured in 17 different countries. There are 7022 1080p\footnote{We also provide the full resolution raw data, as described in Section \ref{addit}.} images in total, captured using three different cameras: Canon 5DSR, Nikon D810, and Mobile Sony IMX135. The dataset has four folders per camera: \textit{field\_1\_camera}, containing unique field images captured by the camera,  \textit{field\_3\_cameras} containing  images of common\footnote{There are 144 scenes that were pictured by the three camera models. For each camera, the folder \textit{field\_3\_cameras}  contains these images of the scenes for the specific camera model.} scenes captured by all cameras, \textit{lab\_printouts}, containing lab printouts, and \textit{lab\_realscenes} consisting of real lab scenes. Table \ref{tablecontent} reports the numbers of images per category.

 \begin{table}[tb]

\renewcommand{\arraystretch}{1}
\footnotesize\setlength{\tabcolsep}{2pt}

 \centering	

	\caption{INTEL-TAU Composition}

	\label{tab:parameters}
	\begin{tabular}{l|cccc} 
		\hline
         &  field\_1\_cameras &  field\_3\_cameras &  lab\_printouts & lab\_realscenes \\	
        \hline
Canon   &  1645 &  144 &  300 & 20  \\	
Nikon  &  2329 &  144 &  300 & 20  \\	
Sony     &  1656 &  144 &  300 & 20  \\	

           \hline
	\end{tabular}
\label{tablecontent}	
\end{table}

When capturing the images, we avoided strong mixed illumination. Instead, we targeted the framing so that one illumination is dominating in the scene. To define the ground truth, there is one ground truth raw Bayer image associated with each raw Bayer image in the database. The ground truth image has a X-Rite ColorChecker Passport chart positioned in such way that it reflects the main illumination in the scene. The actual database image does not contain the chart, except for a handful of images in which it was intentionally inserted as image contents. The same ground truth image can be associated with multiple database images if the illumination is common in those images. We calculated the ground truth white point from grey patches $\#20-\#23$, omitting the brightest grey patch $\#19$, the darkest grey patch $\#24$, and additional saturated patches if any. Noise was reduced by a $9\times9$ averaging filter before recording the color component values inside the center area of the grey patch. We manually checked the annotation  for each image.

\begin{figure}[t]
\centering
\includegraphics[width= 0.5\textwidth]{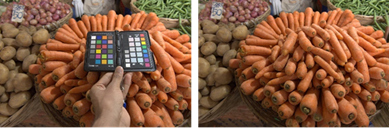}
\caption{An example ground truth and database image pair (illustration, not actual raw Bayer content)}
\label{figg1}
\end{figure}
\begin{figure}[t]
\centering
\includegraphics[width= 0.5\textwidth]{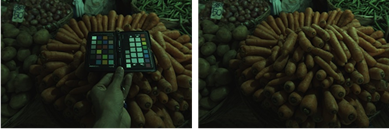}
\caption{An example ground truth and database image pair (actual raw image) }
\label{figg2}
\end{figure}

The associated .ccm was not calculated based on the ground truth image, but selected from a pre-calculated set of CCMs according to the estimate of the illumination (daylight, indoor fluorescent, indoor tungsten-halogen). Consequently, the .ccm should not be treated as an accurate color conversion matrix, but just for more convenient illustration. It can further serve as a means to guide the color shading correction that was applied on the Sony IMX135 images. Figure \ref{figg1} presents an example of ground truth and database image pair as an illustration (this in not the actual raw Bayer content). Figure \ref{figg2} presents the actual raw images of an example ground truth and database image pair as a reminder to the reader that the database has raw Bayer images. Different camera characteristics are presented in Table \ref{car3}.

Only the database images are made publicly available along with the ground truth illumination.  The ground truth images, i.e., images with the color chart, are not published in this version of the dataset. Thus, no color chart masking needs to be done before evaluating color constancy approaches using the proposed INTEL-TAU dataset. In addition, the black level was subtracted, the saturation points were normalized, and all images were down-sampled to 1080p. The images are stored in TIFF format and the associated groundtruths in the normalized [R,G,B] coordinates.  Following the \ac{GDPR} regulations, we applied privacy masking for recognizable faces, license plates, and other privacy sensitive information. The color component values inside the privacy masking area were averaged.

\begin{table*}[!h]
\renewcommand{\arraystretch}{1}
 \centering	
	\caption{Characteristics of the camera models used in Intel-TAU}

	\label{tab:parameters}
	\begin{tabular}{l|ccc} 
		\hline
         &  Canon EOS 5DSR &  Nikon D810 &  Mobile  \\	
        \hline
    Resolution    &  52Mpix (8896H $\times$  5920V) &  36Mpix (7380H $\times$ 4928V) &  8Mpix (3264H $\times$ 2448V)  \\	
    Focal length    &  EF 24-105/4L @ 28mm (*) & AF-S 24-70/2.8G @ 28mm (*)  &  30.4mm (actual 4.12mm)   \\	
    Aperature size   &  F8.0(**) &F8.0(**) &  F2.4   \\	
    Pixel size   &  4.14um & 4.88um &  1.12um  \\	
    raw data bit depth  &  14bpp & 14bpp &  10bpp  \\	
           \hline

\multicolumn{4}{l}{(*)\footnotesize{: 28mm was the closest to the mobile device focal length that was easy to set consistently based on the markings on the objectives}} \\
\multicolumn{4}{l}{(**)\footnotesize{: Smaller aperture was used in order to reduce the depth-of-field difference between the DSLRs and the mobile module}}
	\end{tabular}
	\label{car3}	

\end{table*}

\section{additional resources} \label{addit}
We also provide the original raw version, i.e., raw Bayer, of the dataset\footnote{http://urn.fi/urn:nbn:fi:att:f8b62270-d471-4036-b427-f21bce32b965}.  Different image characteristics are presented in Table \ref{car2}. The white points are stored as [R/G, B/G] coordinates.  The spectral responses of the different camera models and the spectral power distributions of the lab light sources are also provided.  It should be noted that the size of the raw image set is large, 290GB, compared to 50GB of the preprocessed dataset images. This variant of the dataset can be used to study the color shading effect as we provide the mobile images before and after color shading. Since raw Bayer images are provided, this dataset can also be used to study various imaging problems, such as demosaicing \cite{syu2018learning}. The downscaled 1080p version offers a more easily transferrable database size for those use cases that do not require the full resolution, such as color constancy. We also provide a preprocessed downscaled version of the dataset with isotropic scaling: in this variant, the resized images have a height of 1080 lines and the width was adjusted accordingly to preserve the original aspect ratio of the images.

 \begin{table}[h]

\renewcommand{\arraystretch}{1}
\footnotesize\setlength{\tabcolsep}{2pt}

 \centering	

	\caption{Image Characteristics in the original raw images of Intel-TAU}

	\label{tab:parameters}
	\begin{tabular}{l|ccc} 
		\hline
         &  Canon 5DSR &  Nikon D810 &  Sony IMX135  \\	
        \hline
   Image width    &  8896 &  7380 &  3264  \\	
   Light shielded pixels at left   &  160 & 0  &  0   \\	
Image height   &  5920 & 4928 &  2448   \\	
   Light shielded pixels at top   &  64 & 0  &  0   \\	
    Bayer order  &  $RG\_GB$ & $RG\_GB$ &  $GR\_BG$  \\	
        Raw data bit depth(*)  &  14 & 14 &  10  \\	
       Data pedestal/black level  &  2047 & 601 &  64  \\	
       Saturation point (**)  &  15380 & 16383 &  1023  \\	

           \hline

\multicolumn{4}{l}{(*)\footnotesize{: The raw frames are stored as uint16 value per each pixel}} \\
\multicolumn{4}{l}{(**)\footnotesize{: Note that the saturation point is not necessarily $2^{raw\_bpp} -1$ }} \\
\multicolumn{4}{l}{(**)\footnotesize{: Some of the Sony IMX135 images are upside down}}

	\end{tabular}
\label{car2}	
\end{table}

\section{Experimental protocols}
\label{exprotocols}
We propose two experimental protocols for using the proposed INTEL-TAU dataset. The first protocol is for evaluating the camera invariance of the models. In the cross-validation protocol, a 10-fold non-random cross-validation experiment is conducted. 

\subsection{Camera invariance protocol}
The proposed INTEL-TAU dataset can be used to evaluate the camera invariance of color constancy approaches, similarly to \cite{17}.  To this end, all scenes acquired by one camera are used for training, all scenes acquired by a second camera are used for validation, and all scenes acquired by a third camera are used for testing in three experiments:

    \begin{enumerate}
          \item  Images acquired by Canon as a training set, images acquired by Nikon for validation, and Sony images for testing (training: 2109 images, validation: 2793 images, testing: 2120 images).
            \item Images acquired by Nikon for training, images acquired by Sony for validation, and Canon images for testing,
            (training: 2793 images, validation: 2120 images, testing: 2109 images),
            \item Images acquired by Sony for training, images acquired by Canon for validation, and Nikon images for testing
            (training: 2120 images, validation: 2109 images, testing: 2793 images).
    \end{enumerate}

Results are reported as the mean of the results over the three experiments. This test protocol evaluates the camera generalization of the approaches. However, it should be noted that the dataset has multiple samples with similar scene content under various illuminations. Thus, splitting the dataset per camera as illustrated above or using a random split can lead to a partition with a high correlation between the training set  and the test set images. To avoid such scenario, we design a second evaluation protocol which is based on 10-fold non-random cross-validation.

\subsection{Cross-validation protocol }
Similar to other color constancy datasets, INTEL-TAU contains samples from the same geographical location under different illuminations or using different cameras. Thus, random splitting of the dataset might result in a high correlation between the contents of the training and test set images.  To avoid this problem, we propose a non-random 10-fold split of the data  to be used for cross-validation. The proposed subset division is provided along with the dataset. In Table \ref{subsets_car}, we illustrate the characteristics of each subset. We further split the training set, i.e., formed by the nine remaining subsets, by randomly selecting 70\% for  training and 30\% for validation. In total, we have ten experiments and the mean of the achieved results is reported as the final result.

Each subset has around 700 samples except for the first one which has 464 images. Using this evaluation scheme, we have more than 6200 samples for the training and validation in each split. Thus, it is by far the largest training set available for color constancy evaluation. This is extremely useful especially for the evaluation of deep learning-based methods.  The results are reported as the mean value of these ten experiments.

\begin{table}[h]
\caption{ Characteristics of the 10 subsets of the INTEL-TAU used for evaluation with the cross-validation protocol}
\begin{tabular}{c|ccl}
\hline
Subset IDs & Subset size & Camera & Country                  \\ \hline
\hline

01             & 464         & Nikon  & Finland                  \\
02             & 724         & Canon  & India                    \\
03             & 701         & Canon  & Diverse set of countries \\
04             & 684         & Canon  & Diverse set of countries \\
05             & 803         & Nikon  & Tenerife                 \\
06             & 700         & Nikon  & India and Finland        \\
07             & 826         & Nikon  & Iceland and Finland      \\
08             & 645         & Sony   & Malta                    \\
09             & 750         & Sony   & Diverse set of countries \\
10             & 725         & Sony   & Diverse set of countries \\
\hline

\end{tabular}
\label{subsets_car}	
\end{table}

\section{Experimental results} \label{sec:experimentalresults}
For all experiments, we report the mean of the top 25\%, the mean value, the median, Tukey's trimean, and the mean of the worst 25\% of the 'Recovery angular error' \cite{21} between the ground truth white point and the estimated illuminant defined as follows:
\begin{equation}
     \text{$e_{recovery}$}(\textbf{I}^{gt},\textbf{I}^{est})= \cos^{-1} ({ \frac{ \textbf{I}^{gt} \textbf{I}^{est}}{\| \textbf{I}^{gt} \| \|\textbf{I}^{est} \| } }), 
\end{equation}
where $\textbf{I}^{gt}$ is the ground truth illumination  and $\textbf{I}^{est}$ is the estimated illumination. In \cite{finlayson2016reproduction}, another robust metric for evaluating the performance of a color constancy method  called 'Reproduction angular error' was proposed. It is defined as follows:

\begin{equation}
     \text{$e_{reproduction}$}(\textbf{I}^{gt},\textbf{I}^{est})= \cos^{-1} ({ \frac{ \textbf{I}^{gt}/ \textbf{I}^{est} \hspace{2mm} \textbf{w}}{\| \textbf{I}^{gt} /\textbf{I}^{est} \|  \sqrt3} }), 
\end{equation}
where $/$ is the element wise division operator and $\textbf{w}$ is defined as the unit vector, i.e., $\textbf{w} = [1 , 1, 1 ]^T$. We also provide results using this error metric.   

In our experiments, we considered the following static methods:  Grey-World \cite{d4}, White-Patch \cite{funt2000retinex}, Spatial domain \cite{nus}, Shades-of-Grey \cite{d6}, and Weighted Grey-Edge \cite{d8}, Greyness Index 2019  \cite{qian2019finding}, Color Tiger   \cite{banic2017unsupervised}, PCC\_Q2, and the method reported in \cite{yang2015efficient}.  Furthermore, we evaluated the performance of the following learning-based methods: Fast Fourier Color Constancy  (FFCC) \cite{46440},  Fully Convolutional Color Constancy With Confidence-Weighted Pooling (FC$^4$) \cite{44}, Bianco CNN \cite{22}, Color Constancy Convolutional Autoencoder (C3AE)  \cite{mine}, and Bag of Color Features \cite{f1}.

\subsection{Camera invariance protocol}
Table \ref{tab:intelt1} reports the results of several  color constancy approaches using the camera invariance protocol. For the unsupervised approaches, we note high error rates for Grey-world, White-Patch, and the variants of Grey-Edge especially in terms of the mean and the worst 25\%.  Shades of Grey achieves the best results across all metrics for both error functions, $e_{recovery}$ and $e_{reproduction}$.

The supervised approaches yield  lower error rates compared to the unsupervised methods especially in terms of the mean and worst 25\%. For example, in terms of the worst 25\% the top unsupervised method, i.e., Shades-of-Grey, achieves $9\degree$ in $e_{recovery}$ compared to   $7.2\degree$ for the worst supervised method, i.e., Bianco. We note a similar analysis for the $e_{reproduction}$ error metric. For the supervised methods, $FC^4$ achieves the best performance, especially in terms of the median and the worst 25\%.

\begin{table*}[h]
	\caption{Results using INTEL-TAU Dataset using camera invariance protocols}
		\label{tab:intelt1}
 \centering	

\begin{tabular}{l|lllll||lllll}
                    & \multicolumn{5}{c||}{$e_{recovery}$}                                                       & \multicolumn{5}{c}{$e_{reproduction}$}                                                  \\ \hline
Method              & Best \newline 25\% & Mean & Med. & Tri. & W. \newline 25\% & Best \newline 25\% & Mean & Med. & Tri. & W. \newline 25\% \\
\hline
Grey-World   \cite{d4}       &  0.9 &  4.7 &  3.7 & 4.0 & 10.0       &  1.1 &  5.7 &  4.6 & 4.9 & 11.9                      \\
White-Patch   \cite{funt2000retinex}   & 1.1 &  7.0 &  5.4 & 6.2 & 14.6  &  1.3 &  7.5 &  6.3 & 6.7 & 15.7                           \\
Grey-Edge  \cite{d7}      & 1.0 &  5.3 &  4.1 & 4.5 & 11.7 & 1.2 &  6.2 &  4.9 & 5.2 & 13.4 \\
2nd order Grey-Edge \cite{d7} &  1.0 &  5.1 &  3.8 & 4.2 & 11.3 & 1.2 &  6.0 &  4.6 & 4.8 & 13.1  \\
Shades-of-Grey \cite{d6}   &  0.7 &  4.0 &  2.9 & 3.2 & 9.0 & 0.8 &  4.8 &  3.6 & 3.9 & 10.9   \\
Cheng et al. 2014 \cite{nus}  &  0.7 &  4.6 &  3.4 & 3.7 & 10.3  &0.9 &  5.5 &  4.2 & 4.5 & 12.1  \\
Weighted Grey-Edge \cite{d8} &  0.9 &  6.0 &  4.2 & 4.8 & 14.2  &  1.1 &  6.8 &  5.0 & 5.5 & 15.6 \\
\hline
Bianco  \cite{22} &  0.8&  3.4   & 2.5  & 2.7  & 7.2                             &  1.0                            & 4.3 & 3.2  & 3.4  & 9.3 \\
C3AE \cite{mine} &  0.9&  3.4    & 2.7  & 2.8  & 7.0                             &  1.1& 3.9  & 3.3  & 3.5  & 8.8 \\

BoCF \cite{f1} &  0.9&  2.9   & 2.4  & 2.5  & 6.1                           &  0.9    & 3.6& 2.8  &  2.9& 7.5 \\
FC$^4$ (VGG16) \cite{44} &  0.7&  2.6    & 2.0  & 2.2  & 5.5                             &  0.8                             & 3.3  & 2.6  & 2.7  & 7.1 \\

\end{tabular}
\end{table*}

\subsection{Cross-validation protocol}

\begin{figure}[t]
\centering
\includegraphics[width= 0.5\textwidth]{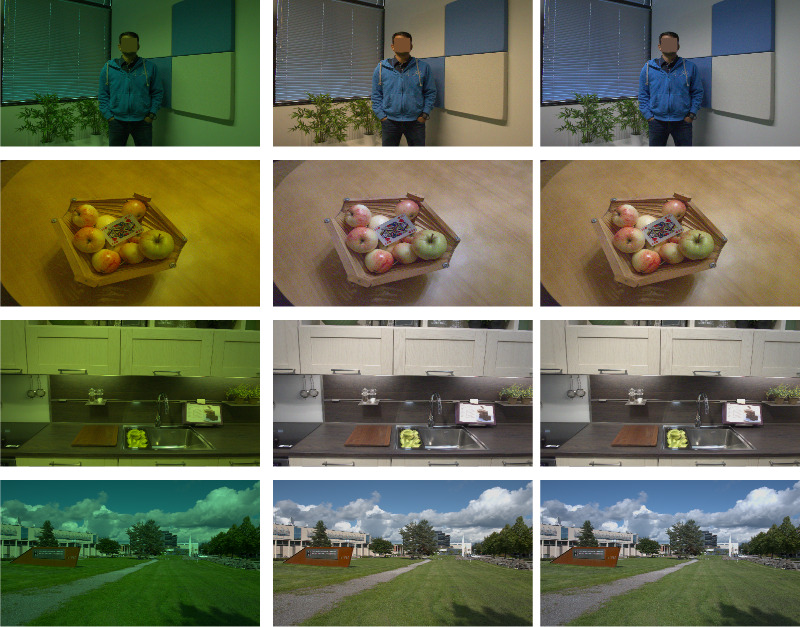}
\caption{Visual  results  on  INTEL-TAU  using BoCF method \cite{f1}. From left to right,  our input images from INTEL-TAU, our corrected images with BoCF method \cite{f1}, and the ground truth image.}
\label{test}
\end{figure}

We perform more extensive experiments using the cross-validation protocol. Table \ref{tab:intelt} reports the results of different color constancy techniques. For the unsupervised approaches, we note high errors for both angular error metrics. The method in \cite{yang2015efficient}  achieves the best results across all the metrics except for the best $25\%$ using $e_{recovery}$, where Grayness Index achieves the smallest errors. It should be noted that the worst 25\% error rate is high for all these approaches (larger than 7\degree).

We note that the supervised methods largely outperform  most of the unsupervised approaches, especially in terms of the worst 25\%.   FFCC, despite not being a deep learning-based approach, achieves competitive results compared to the \acf{CNN} methods. This can be noted especially in terms of the trimean and median. FC$^4$ achieves the lowest error rates across all metrics expect for the median, where FFCC achieves the best results.

Figure \ref{test} illustrates visual results on three  INTEL-TAU image  samples, along with the output of BoCF \cite{f1} and the associated ground truth.  We note that for a different types of scenes, the model manages to recover the original colors of the scene and produce visually similar results to the output.\\

In Table \ref{tab:final}, we report the median result of various CNN-based approaches, i.e., FC$^4$, Bianco, C3AE, and BoCF on the different folds of the split. We note high error rates for specific folds: the first, forth, and seventh subsets. We also note that the second, and ninth sets are easy to learn and most CNN-based approaches achieve less than  2.3\degree median error.

\begin{table*}[h]
	\caption{Results of benchmark methods on INTEL-TAU Dataset using cross-validation protocol.}
		\label{tab:intelt}
 \centering	

\begin{tabular}{l|lllll||lllll}
                    & \multicolumn{5}{c||}{$e_{recovery}$}                                                       & \multicolumn{5}{c}{$e_{reproduction}$}                                                  \\ \hline
Method              & Best \newline 25\% & Mean & Med. & Tri. & W. \newline 25\% & Best \newline 25\% & Mean & Med. & Tri. & W. \newline 25\% \\
\hline
Grey-World  \cite{d4}        & 1.0                               & 4.9  & 3.9  & 4.1  & 10.5                            & 1.2                               & 6.1  & 4.9  & 5.2  & 13.0                            \\
White-Patch \cite{funt2000retinex}     & 1.4                               & 9.4  & 9.1  & 9.2  & 17.6                            & 1.8                               & 10.0  & 9.5  & 9.8  & 19.2                            \\
Grey-Edge   \cite{d7}    & 1.0                               & 5.9  & 4.0  & 4.6  & 13.8                            & 1.2                               & 6.8  & 4.9  & 5.5  & 13.5                           \\
2nd order Grey-Edge \cite{d7} & 1.0                               & 6.0  & 3.9  & 4.8  & 14.0                            & 1.2                               & 6.9  & 4.9  & 5.6  & 15.7                            \\
Shades-of-Grey \cite{d6}  & 0.9                               & 5.2  & 3.8  & 4.3  & 11.9                             & 1.1                               & 6.3  & 4.7  & 5.1  & 13.9                            \\
Cheng et al. 2014 \cite{nus} & 0.7 & 4.5  & 3.2  & 3.5  & 10.6                            &  0.9                               & 5.5  & 4.0  & 4.4  & 12.7                            \\
Weighted Grey-Edge  \cite{d8} &  0.8                               & 6.1  & 3.7  & 4.6  & 15.1                            & 1.1                               & 6.9  & 4.5  & 5.4  & 16.5                            \\
Yang et al. 2015  \cite{yang2015efficient} &  0.6                               &  3.2  &  2.2  &  2.4  &  7.6                             &  0.7                               &  4.1  &  2.7  &  3.1  &  9.6                            \\
Color Tiger   \cite{banic2017unsupervised}        & 1.0     & 4.2    & 2.6  & 3.2  & 9.9    & 1.1      & 5.3  &  3.3  & 4.1  & 12.7   \\
Greyness Index  \cite{qian2019finding} &  0.5                               &  3.9    &  2.3  &  2.7  &  9.8                             &  0.6                               &  4.9  &  3.0  &  3.5  &  12.1   \\

PCC\_Q2 \cite{laakom2020probabilistic} &  0.6                               &  3.9    &  2.4  &  2.8  &  9.6                             &  0.7 &  5.1& 3.5  &  4.0  &  11.9  \\
\hline

FFCC\cite{{46440}} &  0.4                              &  2.4    & 1.6  & 1.8  & 5.6                             &  0.5                             & 3.0  & 2.1  & 2.3  & 7.1 \\

Bianco  \cite{22} &  0.9&  3.5   & 2.6  & 2.8  & 7.4                             &  1.1                            & 4.4 & 3.4  & 3.6  & 9.4 \\
C3AE \cite{mine} &  0.9&  3.4    & 2.7  & 2.8  & 7.0                             &  1.1& 3.9  & 3.3  & 3.5  & 8.8 \\

BoCF \cite{f1} &  0.7&  2.4   & 1.9  & 2.0  & 5.1                           &  0.8    & 3.0& 2.3  &  2.5& 6.5 \\
FC$^4$ (VGG16) \cite{44} &  0.6&  2.2    & 1.7  & 1.8  & 4.7                             &  0.7                             & 2.9  & 2.2  & 2.3  & 6.1 \\
\end{tabular}
\end{table*}

\begin{table*}[h]
	\caption{Median errors of $e_{recovery}$ per split for CNN-based approaches for the cross-validation protocol.}
		\label{tab:final}
 \centering	

\begin{tabular}{l|llllllllll}
                    & \multicolumn{10}{c}{ Subset IDS }                                                  \\ \hline
Method              & 01 & 02 & 03 & 04 & 05 & 06 & 07 & 08 & 09 & 10 \\
\hline

Bianco  \cite{22} &  3.6&  2.2   & 2.2  & 4.1  & 1.2                             &  2.3                             & 3.0& 2.8  & 2.0  & 2.5 \\

C3AE \cite{mine} &  3.9&  2.3  & 2.6  & 3.0& 2.2  & 2.5 &  3.0    & 2.5  & 2.3  & 2.6  \\

BoCF \cite{f1} &  2.3&  1.7    & 1.6  & 1.8  & 1.6&  1.8    & 2.2& 1.7 &  2.0& 2.0 \\ 
FC$^4$ (VGG16) \cite{44}&  2.0&  1.5   & 1.5  & 1.8  & 1.7 &  1.8 & 2.0& 1.5  &  1.7& 1.9 \\

\end{tabular}
\end{table*}

\section{Conclusion} \label{sec:conclusion}

In this paper, a new  color constancy dataset, namely INTEL-TAU, is presented. It is the largest available dataset and thus the most suitable for deep learning methods evaluation. The diversity of scenes and camera models makes the new database appropriate for  evaluating the camera and scene invariance of different illumination estimation techniques. Privacy masking has been applied for sensitive information, e.g., faces, thus, rendering the dataset compliant with the new \ac{GDPR} regulations. Furthermore, the effect of color shading for mobile images can be evaluated with INTEL-TAU, as it provides both corrected and uncorrected versions of the raw mobile data.

\ifCLASSOPTIONcompsoc

\section*{Acknowledgments}
\else
\section*{Acknowledgment}
\fi

This  work  was  supported  by   NSF-Business Finland  Center  for Visual and Decision Informatics (CVDI) project AMALIA 2019. We thank Harish Essaky Sankaran, Uygar Tuna, and Lauri Suomela  for participating in capturing the new images.

\ifCLASSOPTIONcaptionsoff
  \newpage
\fi



%

\bibliographystyle{IEEEtran}

\bibliography{strings}




%

 \vskip -1\baselineskip plus -1fil

\begin{IEEEbiographynophoto}{Firas Laakom}
is a doctoral student at Tampere university,Finland. He received his engineering degree from Tunisia Polytechnic School (TPS) in 2018.  His research interests include deep learning, computer vision
and computational intelligence.
\end{IEEEbiographynophoto}
 \vskip -1\baselineskip plus -1fil
\begin{IEEEbiographynophoto}{Jenni Raitoharju} 
received her Ph.D. degree at Tampere University of Technology, Finland in 2017. Since then, she has worked as a Postdoctoral Research Fellow at the Faculty of Information Technology and Communication Sciences, Tampere University, Finland. In 2019, she started working as a Senior Research Scientist at the Finnish Environment Institute, Jyväskylä, Finland after receiving Academy of Finland Postdoctoral Researcher funding for 2019-2022. She has co-authored 13 journal papers and 27 papers in international conferences. She is the chair of Young Academy Finland 2019-2020. Her research interests include machine learning and pattern recognition methods along with applications in biomonitoring and autonomous systems.\end{IEEEbiographynophoto}
 \vskip -1\baselineskip plus -1fil
\begin{IEEEbiographynophoto}{Jarno Nikkanen}
received his M.Sc. and Dr.Sc.Tech. degrees from Tampere University of Technology in 2001 and 2013, respectively, with subjects in Signal Processing and Software Systems. Jarno has 18 years of industry experience in digital imaging topics, starting at Nokia Corporation in 2000 where he developed and productized many digital camera algorithms, and moving to Intel Corporation in 2011 where he is currently working as Intel Principal Engineer and Imaging Technology Architect. Jarno holds international patents for over 20 digital camera related inventions.\end{IEEEbiographynophoto}
 \vskip -1\baselineskip plus -1fil
\begin{IEEEbiographynophoto}{Alexandros Iosifidis}
is an Associate Professor at Aarhus University, Denmark. He has contributed in more than ten R\&D projects financed by EU, Greek, Finnish, and Danish funding agencies and companies. He has co-authored 53 articles in international journals and 78 papers in international conferences proposing novel Machine Learning techniques and their application in a variety of problems. Dr. Iosifidis is a Senior Member of IEEE and he served as an Officer of the Finnish IEEE Signal Processing-Circuits and Systems Chapter.\end{IEEEbiographynophoto}
 \vskip -1\baselineskip plus -1fil
\begin{IEEEbiographynophoto}{Moncef Gabbouj}
received his MS and PhD degrees in EE from Purdue University, in 1986 and 1989, respectively. Dr. Gabbouj is Professor of Signal Processing at the Department of Computing Sciences, Tampere University, Finland. His research interests include Big Data analytics, multimedia analysis, artificial intelligence, machine learning, pattern recognition, nonlinear signal processing, video processing and coding. Dr. Gabbouj is a Fellow of the IEEE and member of the Academia Europaea and the Finnish Academy of Science and Letters.\end{IEEEbiographynophoto}




\end{document}